\documentstyle[11pt,paspconf,epsf]{article}
\begin{document}
\title{The GL bibliography and an interactive database}
\author{A. Pospieszalska-Surdej, J. Surdej\altaffilmark{1}, A. Detal 
and C. Jean \affil{Institut d'Astrophysique et de G\'eophysique, Universit\'e 
de Li\`ege \,
Avenue de Cointe 5, B--4000 Li\`ege, Belgium}}
\altaffiltext{1}{Also Directeur de Recherches du Fonds National de la
Recherche Scientifique, Belgium}

\begin{abstract}
It is now possible to directly access, via the Internet, a bibliographical
database on Gravitational Lensing (GL) literature. The Interactive 
Gravitational 
Lensing Bibliography (IGLB) totalizes more than 2400 titles of published 
articles in scientific journals and  meeting proceedings (except those 
fully dedicated to Gravitational Lenses) as well as papers submitted to 
the e-Print archive. This database is a product from the Gravitational Lensing 
Bibliography first presented in 1993 (Proceedings of the 31st Li\`ege 
International Astrophysical Colloquium). It is easy to do field based 
searches for title keywords, authors (using boolean operators), year and 
journal (a pull-down list of the most cited journals is available). Access
to the original version of published articles as well as to preprints 
submitted to the e-Print archive at the URL address http://xxx.lanl.gov/ is 
also provided. This database 
is updated approximately every two months. The "complete" bibliography of 
published articles is also available in the form of Latex and PostScript 
files. The IGLB can be accessed at the URL: 
http://vela.astro.ulg.ac.be/grav\_lens
\end{abstract}


\section{Introduction}
Figure 1 shows the WWW homepage for the gravitational lensing database.
By clicking on one of the four images of the Clover-leaf, one may have access 
to different entries of the gravitational lensing bibliography:

- 1) a Database with field-based searches totalizing more than 2400 articles 
published and submitted to main scientific journals with corresponding links 
(for submitted papers to the e-Print archive and for published papers to 
main journals),

- 2) a PostScript file containing published articles,

- 3) a Latex file also containing published articles,

- 4) a Books file containing conference proceedings and books fully dedicated 
to GL. 

 The first version of the GL bibliography approximately totalized 1000 
articles. The rapid growth of the GL activity is shown in Figure 2 which 
illustrates the number of GL related articles being published each year.

\newpage
\vspace*{2cm}
\begin{figure}[ht]
\plotfiddle{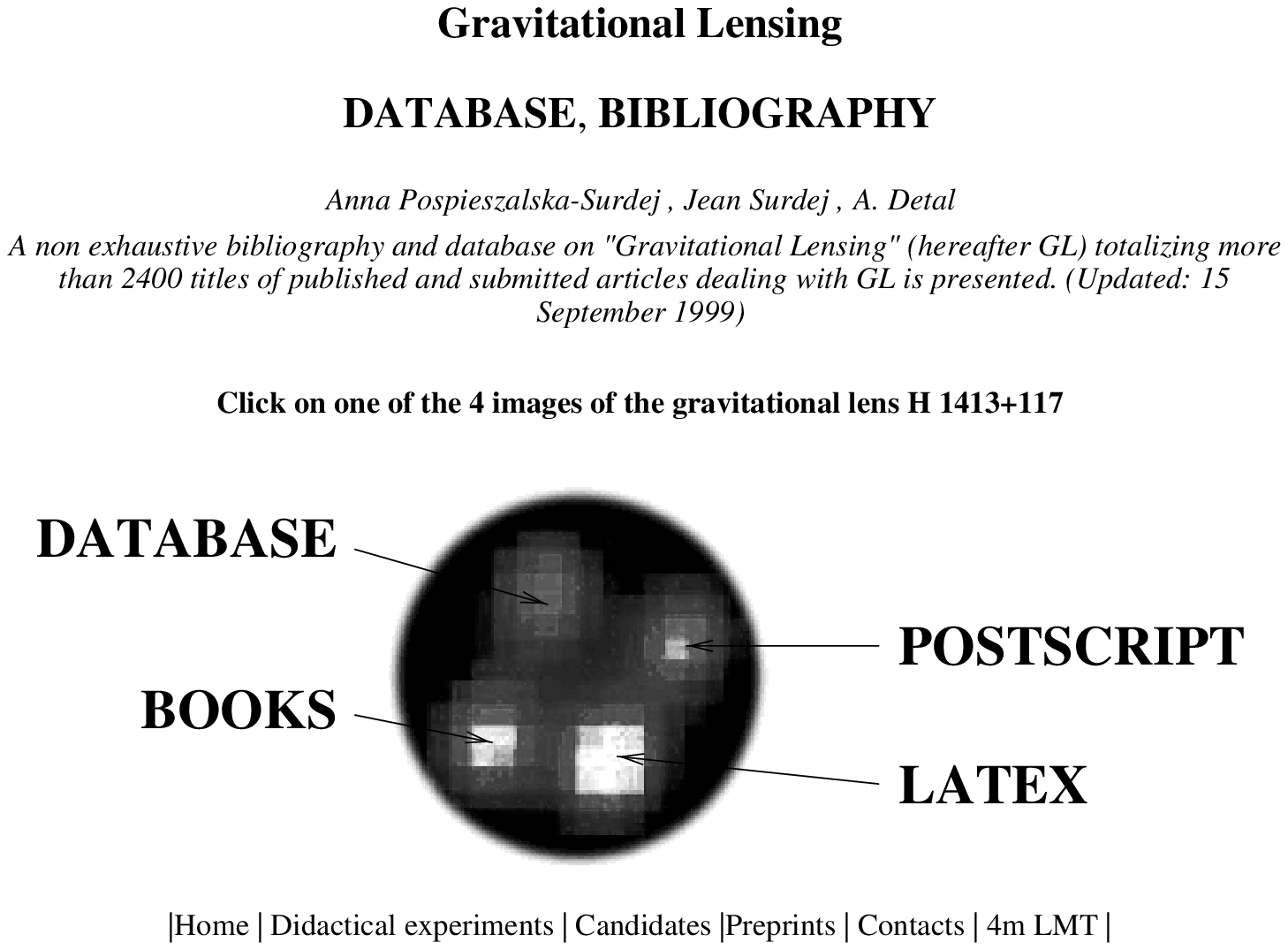}{5cm}{0}{75}{75}{-190}{-250}
\label{fig-1}
\caption{World Wide Web homepage for the GL database and bibliography 
accessible at the URL http://vela.astro.ulg.ac.be/grav\_lens.} 
\end{figure}

\begin{figure}[h]
\plotfiddle{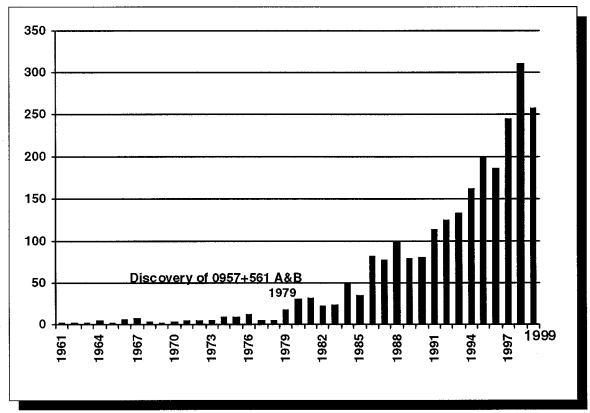}{5cm}{0}{60}{60}{-110}{-5}
\label{fig-2}
\caption{Number of papers related to gravitational lensing, published per 
year, during the past forty years. Please note that the year 1999 is not
yet over!} 
\end{figure}


\begin{references} 
\reference Pospieszalska-Surdej, A., Surdej, J., V\'eron, P. 1993, 
Proceedings of the 31st Li\`ege International Astrophysical Colloquium, p. 671
\reference Pospieszalska-Surdej, A., Surdej, J., V\'eron, P. 1995, Proceedings 
of the International Astrophysical Union Symposium No. 173, p. 417
\end{references}
\end{document}